\documentstyle[12pt]{article}

\newcommand{\resection}[1]{\setcounter{equation}{0}\section{#1}}
\newcommand{\appsection}{\addtocounter{section}{1} \setcounter{equation}{0}
                         \section*{Appendix \Alph{section}}}

\newcommand{\EQ}{\begin{equation}}
\newcommand{\EN}{\end{equation}}
\newcommand{\bea}{\begin{eqnarray}}
\newcommand{\eea}{\end{eqnarray}}
\newcommand{\hs}{\hspace{0.1cm}}

\newcommand{\s}{\sigma}
\newcommand{\goto}{\rightarrow}
\newcommand{\be}{\beta}

\begin{document}
\setcounter{page}{0}
\topmargin 0pt
\oddsidemargin 5mm
\newpage
\setcounter{page}{0}
\begin{titlepage}
\begin{flushright}
ISAS/EP/93/68\\
\end{flushright}
\vspace{0.5cm}
\begin{center}
{\large {\bf Stress-Energy Tensor and Ultraviolet Behaviour in
Massive Integrable Quantum Field Theories}}
\\
\vspace{1.8cm}
{\large G. Mussardo and P. Simonetti}\\
\vspace{0.5cm}
{\em International School for Advanced Studies,\\
and \\
Istituto Nazionale di Fisica Nucleare\\
34014 Trieste, Italy}\\
\end{center}
\vspace{1.2cm}
\renewcommand{\thefootnote}{\arabic{footnote}}
\setcounter{footnote}{0}
\begin{abstract}
\noindent
The short distance behaviour of massive integrable quantum field theories
is analyzed in terms of the form factor approach. We show that the on-shell
dynamics is compatible with different definitions of the stress-energy tensor
$T_{\mu\nu}(x)$ of the theory. In terms of form factors, this is equivalent
to having a possible non-zero matrix element $F_1$ of the trace of $T_{\mu\nu}$
on one-particle state. Each choice of $F_1$ induces a different scaling
behaviour of the massive theory in the ultraviolet limit.
\end{abstract}
\end{titlepage}
\newpage
\resection{Introduction}

One of the most relevant steps towards the physical interpretation of a given
field dynamics consists in the identification of the stress tensor
$T_{\mu\nu}(x)$. It gives the local distribution of energy and momentum, and
rules the response of the system under local scale transformations.
In this paper we consider quantum field theories involving one scalar
field $\varphi(x)$ defined in ($1+1$) flat space-time where there is, however,
a natural ambiguity in the definition of $T_{\mu\nu}(x)$. To see this, the
simplest way is to initially consider the theory defined on a curved
two-dimensional manifold with metric tensor $g_{\mu\nu}$ and
scalar curvature $R$
\EQ
{\cal A}\,=\,\int\,d^2\xi\,\sqrt{- g}
\left(\frac{1}{2}g^{\mu\nu}\,\partial_{\mu}\varphi\partial_{\nu}\varphi\,-\,
V(\varphi)+\alpha\,R\,\varphi\right)\,\,\, ,
\EN
In the above action, $\alpha$ is a free parameter that in the limit of a
flat manifold $g_{\mu\nu}\rightarrow \eta_{\mu\nu}$, labels the
one-dimensional family of stress-energy tensors associated to the on-shell
dynamics
\EQ
T_{\mu \nu}(x)\,=\,\tilde T_{\mu \nu}(x)
\,+\, \alpha \left( \partial_{\mu} \partial_{\nu}
\,-\, g_{\mu \nu} \Box \right) \varphi(x)
\,\,\, ,
\label{alpha}
\EN
where
\EQ
\tilde T_{\mu\nu}(x)\,=\,2\pi\left[
\partial_{\mu}\varphi\,\partial_{\nu}\varphi\,-\,
\eta_{\mu\nu}\left(\frac{1}{2}(\partial\varphi)^2\,-\,V(\varphi)\right)
\right]\,\,\,.
\EN
Although in the flat Minkowski space the equations of motion for the field
$\varphi$ do not depend on the specific definition of $T_{\mu\nu}(x)$,
the ultraviolet behaviour of the theory may be extremely sensitive to
any modification of this operator. A well-known example is provided by
the conformal invariant QFT \cite{BPZ,ISZ}: in a Coulomb Gas language, an
extra derivative term in the definition of the stress-energy
tensor results in a non-zero background charge that, in turn, induces a
non-trivial scaling behaviour of the operators of the theory \cite{DF,Felder}.
It is therefore an interesting question to see what are the consequences of a
redefinition of $T_{\mu\nu}(x)$ also in the case of quantum field theories
which do not have scaling invariant properties. A special class of these
theories are the massive integrable models where the off-shell dynamics may be
completely characterized in terms of the on-shell scattering amplitudes of the
massive excitations. In fact, we can take advantage of their integrability, and
compute exactly the matrix elements of local operators ${\cal O}_k(x)$ on the
asymptotic states, by means of the {\em Form Factor Bootstrap Approach} [5-13].
All correlation functions are then reconstructed in terms of their spectral
representation. For large values of the relative distances, these correlation
functions will have an exponential decay ruled by the lowest massive state
appearing in their spectral decomposition. Their short distance behaviour,
on the contrary, present power law singularities. In order to identify the
conformal dimensions of the operators, we need to analyze the short distance
singularity of the correlation functions
$<T_{zz}(z,\overline z)\,{\cal O}_k(0)>$. For those operators that correspond
to primary fields in the conformal limit $(mR)\goto 0$
($R^2=z\overline z$), we have
\EQ
<T_{zz}(z,\overline z)\,{\cal O}_k(0)> \,\simeq\,\frac{\Delta({\cal O}_k)}{z^2}
\,\,\,.
\EN
The conformal dimensions $\Delta({\cal O}_k)$ will depend in general on the
definition of $T_{\mu\nu}(x)$ and therefore different definitions of
this operator may induce different scaling behaviours of the theory in
the deep ultraviolet region. Analogously, the central charge of the ultraviolet
theory may be extracted from the short-distance singularity of the correlator
\EQ
<T_{zz}(z,\overline z)\,T_{zz}(0)> \,\simeq\,\frac{c}{2 z^4}
\,\,\,.
\EN

The plan of the paper is as follows. In section 2 we discuss the general
properties of the stress-energy tensor. Using entirely the formalism of
the form factor approach, we derive the conditions on the matrix elements of
$T_{\mu\nu}(x)$ which guarantee its conservation and locality. From this
analysis, there is a one-dimensional space of possible stress-energy
tensors $T_{\mu\nu}(x)$ compatible with the on-shell dynamics. In section 3 and
4 we then discuss the ultraviolet behaviours of the simplest Affine
Toda Field Theories \cite{MOP}, i.e. the Sinh-Gordon and the Bullough-Dodd
models based respectively on the simply-laced algebra $A_1^{(1)}$ and on
the non-simply laced algebra $A_2^{(2)}$. In both theories, different
ultraviolet limits may be reached by varying the definition of the
stress-energy tensor, with the relevant conformal data simply obtained in
terms of the form factors. Our conclusions are summarized in section 5.

\resection{Form Factors of the Stress-Energy Tensor}

Let us consider an integrable two-dimensional massive QFT characterized by
its elastic factorizable $S$-matrix \cite{ZZ,Zam,Musrep}. As it is well known,
the form factor approach is quite effective to characterize the operators in
such a theory in terms of their matrix elements on the set of asymptotic states
\cite{KW,Smirnov}. Adopting the standard parametrization of the momenta in
terms of the rapidity variable
$\beta$, i.e. $p^{\mu}\,=\,(m\,\cosh \be,\,m\,\sinh \be)$, any local operator
${\cal O}_k(x)$ will be uniquely identified\footnote{This justifies the
interchangeable use we make of the words ``operator'' or ``form factors''
in the rest of the paper.} by the set of its Form Factors (FF)
\EQ
F^k_n(\be_,\ldots,\be_n)\,=\,<0|{\cal O}_k(0)|\be_1, \ldots, \be_n>_{\rm in}.
\EN
In the above definition, the set $\{\be_i\}$ is ordered as
$\be_1\,>\beta_2\,\ldots \,>\be_n$ and our normalization is given by
\EQ
{}_{\rm in}<\be'_1, \,\ldots,\,\be'_m\,|\,
\be_1,\,\ldots,\,\be_n>_{\rm in}\,=\, \delta_{m,n}\,
\prod_{i=1}^n\,2\pi\,\delta (\be'_i\,-\,\be_i)
\,\,\,.
\EN
Once the FF of the local operators ${\cal O}_k(x)$ are known, their two-point
(and higher) correlation functions can be reconstructed through the unitary
sum\footnote{This expression holds for scalar operators. The generalization
of this equation for operators of spin $s$ is easily done by using
eq.\,(\ref{asymp2}).}
\EQ
<{\cal O}_k(r)\,{\cal O}_k(0)>\,=\,\sum_{n=0}^{\infty}
\int \frac{d\beta_1\ldots d\beta_n}{n! (2\pi)^n}
\mid F_n^k(\beta_1\ldots \beta_n)\mid^2 \exp \left(-mr\sum_{i=1}^n\cosh\beta_i
\right)
\EN
Let us discuss then the properties of the FF, as dictated by relativistic
invariance and general requirements of QFT \cite{KW,Smirnov}. For an operator
${\cal O}_k(x)$ of spin $s$, relativistic invariance implies
\EQ
F_n^k (\beta_1+\Lambda,\beta_2+\Lambda,\ldots,\beta_n+\Lambda) \,=\,
e^{s\Lambda}\,
F_n^k (\beta_1,\beta_2,\ldots,\beta_n) \,\, .\label{relat}
\label{asymp2}
\EN

The FF of a given theory are solutions of a set of functional and recursive
equations. The functional equations arise from the monodromy properties of the
functions $F_n^k$ which are ruled by the $S$-matrix\footnote{For simplicity
we consider theory with only one massive self-conjugate particle.}
\bea
F_n^k(\be_1,\dots,\be_i,\be_{i+1},\dots, \be_n) &=& F^k_n(\be_1,\dots,\be_{i+1}
,\be_i ,\dots, \be_n) \,S(\beta_i-\beta_{i+1}) \,\, ,
\label{permu1}\\
F^k_n(\be_1+2\pi i,\dots,\be_{n-1},\be_n ) &=& F^k_n(\be_2 ,\ldots,\be_n,
\be_1) = \prod_{i=2}^{n} S(\beta_i-\beta_1) \,F^k_n(\be_1, \dots, \be_n)
\,\, .
\nonumber
\eea
The recursive equations are obtained, on the contrary, by looking at the pole
structure of the matrix elements $F_n^k$. The first kind of poles are
kinematical poles located at $\beta_{ij}=i\pi$. The corresponding residues
give rise to a recursive equation between the $n$-particle and the
$(n+2)$-particle FF
\EQ
-i\lim_{\tilde\beta \rightarrow \beta}
(\tilde\beta - \beta)
F_{n+2}^k(\tilde\beta+i\pi,\beta,\beta_1,\beta_2,\ldots,\beta_n)=
\left(1-\prod_{i=1}^n S(\beta-\beta_i)\right)\,
F_n^k(\beta_1,\ldots,\beta_n)  \,\,\, .
\label{recursive}
\EN
If bound states are present in the spectrum, there is another set
of recursive equations obtained by looking at their poles in the
matrix elements. Let $\beta_{ij}=i\, u$ be the location of the pole in the
two-particle scattering amplitude corresponding to the bound state. Then the
corresponding residue in the FF is given by
\EQ
-i\lim_{\epsilon\rightarrow 0} \epsilon\,
F_{n+1}(\beta+i\,u-\frac{\epsilon}{2},
\beta+\frac{\epsilon}{2},\beta_1,\ldots,\beta_{n-1})
\,=\,\Gamma\,F_{n}(\beta,\beta_1,\ldots,\beta_{n-1})
\,\,\, ,
\label{respole}
\EN
where $\Gamma$ is the on-shell three-particle vertex. This equation
establishes a recursive structure between the $(n+1)$- and $n$-particle form
factors.

The two chains of recursive equations,
\EQ
\begin{array}{l}
\dots \goto \,F_{n+4}\,\goto\,F_{n+2}\,\goto\,F_{n}\,\goto\,F_{n-2}
\,\goto\,\dots\\
\dots \goto \,F_{n+4}\,\goto\,F_{n+3}\,\goto\,F_{n+2}\,\goto\,F_{n+1}
\,\goto\,\dots
\end{array}
\label{doublechain}
\EN
(and the consistency conditions associated to them) are quite effective for
the explicit determination of the FF of a given theory.

The above discussion holds for any FF of a local operator, in particular
for those of the stress-energy tensor $T_{\mu\nu}(x)$. However, due to the
special role of this operator, its FF has some distinguishing properties.
Since $T_{\mu\nu}(x)$ is conserved, it may be expressed in terms of an
auxiliary scalar field $A(x)$ as \cite{Smirnov}
\EQ
T_{\mu \nu}(x)\,=\, \left( \partial_{\mu} \partial_{\nu}
\,-\, g_{\mu \nu} \Box \right) \,A(x)
\,\,\,.
\EN
In light-cone coordinates $x^{\pm}=x^0 \pm x^1$, its components are given by
\EQ
T_{+ +}\,=\,\partial_+^2 \, A\,\,\,\,\,\,\,\,,
T_{- -}\,=\,\partial_-^2 \, A\,\,,
\EN
\EQ
\Theta\,=\,T^{\mu}_{\mu}\,=\,-\,\Box\,A
\,=\,-\,4\,\partial_+ \partial_- \,A
\,\,\,.
\EN
Introducing the variables $x_j\,=\,e^{\be_j}$ and the elementary
symmetric polynomials in $n$-variables $\s^{(n)}_i$ defined by the
generating function
\EQ
\prod_{j=1}^n\,(x\,+\,x_j)\,=\,
\sum_{i=0}^n x^{k-i}\s_i^{(n)}(x_1,\ldots,x_k)\,\,\, ,
\EN
it is easy to see that
\bea
F_n^{T_{++}}(\beta_1,\ldots,\beta_n) &\,=\,& -\,\frac{1}{4}\,m^2\,
\left(\frac{\s_{n-1}^{(n)}}{\s_n^{(n)}} \right)^2\,F_n^{A}
(\beta_1,\ldots,\beta_n) \,\,\, ,
\nonumber \\
F_n^{T_{--}}(\beta_1,\ldots,\beta_n) &\,=\,& -\,\frac{1}{4}\,m^2\,
\left( \s_1^{(n)} \right)^2\,F_n^{A}(\beta_1,\ldots,\beta_n)
\label{auxil} \,\,\, ,\\
F_n^{\Theta}(\beta_1,\ldots,\beta_n) &\,=\,& m^2\,
\frac{\s_1^{(n)} \s_{n-1}^{(n)}}{\s_k} \,F_n^{A}(\beta_1,\ldots,\beta_n)
\,\,\, .
\nonumber
\eea
Solving for $F_n^A$, we have
\bea
F_n^{T_{++}}(\beta_1,\ldots,\beta_n) &\,=\,& -\,\frac{1}{4}\,
\frac{\s_{n-1}^{(n)}}{\s_1^{(n)} \s_n^{(n)}} \,F_n^{\Theta}
(\beta_1,\ldots,\beta_n) \,\,\, ,
\nonumber \\
F_n^{T_{--}}(\beta_1,\ldots,\beta_n) &\,=\,& -\,\frac{1}{4}\,
\frac{\s_1^{(n)} \s_n^{(n)}}{\s_{n-1}^{(n)}} \,F_n^{\Theta}
(\beta_1,\ldots,\beta_n)
\,\,\,.
\eea
Hence the complete knowledge of $T_{\mu \nu}$ is encoded into the form
factors of the trace $\Theta$. As any spinless operator, its form factors
$F_n^{\Theta}(\beta_1,\ldots,\beta_n)$ depend only on the difference of the
rapidities $\be_{ij}\,=\,\be_i\,-\,\be_j$. Moreover, since the FF of
$T_{--}$ and $T_{++}$ have the same singularity structure of the FF of
$\Theta$, $F_n^{\Theta}(\beta_1,\ldots,\beta_n)$ (for $n>2$) has to be
proportional to the combination of symmetric polynomials
$\sigma_1^{(n)}\sigma_{n-1}^{(n)}$ which corresponds to the invariant total
energy-momentum.

Additional constraints on $F_n^{\Theta}$ are obtained from the knowledge of
their asymptotic behaviour in each variable $\beta_i$. This behaviour
generally depends on the particular model under consideration. For the
lagrangian quantum field theories discussed in this paper, we have
\EQ
F_n^{\Theta}(\be_1 + \Delta,\be_2,\ldots,\be_n)\,\,
\stackrel{\Delta \goto \infty}{\longrightarrow}\,\,o(1)
\,\,\, ,
\label{asympclass}
\EN
i.e. they become constant for large values of the individual momenta. This
condition can be easily checked by analyzing the asymptotic behaviour of the
Feynman diagrams entering the perturbative definition of these matrix elements
\cite{Smirnov,FMS}.

Concerning their normalization, the recursive structure of the space of FF
reduces the problem of finding the normalization of the matrix element of
$\Theta(x)$ for the initial conditions of the double chain (\ref{doublechain}),
i.e. the two-particle FF $F_{2}^{\Theta}(\beta_{12})$ and the one-particle
FF $F_1^{\Theta}(\beta)$.

The normalization of the two-particle FF $F_2^{\Theta}(\beta_{12})$ can be
fixed by making use of the definition of the energy operator
\cite{Smirnov}
\EQ
E\,=\,\frac{1}{2 \pi} \int_{-\infty}^{+\infty}\,d x^1\, T^{00}(x)
\,\,\,.
\label{energy}
\EN
In fact, computing the matrix element of both terms of this equation on the
asymptotic states $<\be'|$ and $|\be>$, for the left hand side we have
\EQ
<\be'|\,E\,|\be>\,=\,2 \pi \,m\,\cosh \be\,\delta(\be'-\be)
\,\,\,.
\EN
On the other hand, taking into account that $T^{00}=\partial_1^2 A$ and
using the relation
\EQ
<\be'|\,{\cal O}(x)|\be>\,=\,
e^{i \left(p^{\mu}(\be')\,-\,p^{\mu}(\be) \right)\,x_{\mu} }\,
F_2^{\cal O}(\be,\,\be'\,-\,i \pi)
\EN
valid for any hermitian operator ${\cal O}$, we obtain
\EQ
F_2^{\partial_1^2 A}(\be_1,\be_2)\,=\,
-\,m^2\,(\sinh\be_1\,+\sinh\be_2)^2\,F_2^A(\be_{12})\,\,\, .
\EN
Then, from eqs.\,(\ref{auxil}) and (\ref{energy}), the normalization of
$F_2^{\Theta}$ is given by
\EQ
F_2^{\Theta}(i\pi)\,=\,2\,\pi\,m^2
\,\,\,.
\EN
However, no special constraint exists for the matrix element of $\Theta(x)$
on the one-particle state
\EQ
F^{\Theta}_1\,=\,<0\mid\Theta(0)\mid\beta>\,\,\, ,
\EN
which is then a free parameter of the theory. Notice that from Lorentz
invariance, it does not depend on the rapidity variable $\beta$.

Since higher FF of $\Theta$ are obtained as solutions of the recursive
equations (\ref{recursive}) and (\ref{respole}) (with initial condition
given by the one-particle and two-particle matrix elements), the arbitrariness
of $F_1^{\Theta}$ propagates in the recursive structure (\ref{doublechain}) of
the FF and therefore gives rise to a one-parameter family of possible
stress-energy tensor $T_{\mu\nu}$ for a given theory.

A simple example of the above discussion is provided by the free massive
bosonic theory, with equation of motion
\EQ
(\Box + m^2)\, \varphi \,=\,0 \,\,\, .
\EN
The $S$-matrix in this case is simply $S=1$ and therefore the FF have trivial
monodromy properties and simple analytic structure. Among the local operators
of the theory, the elementary field $\varphi(x)$ is identified by the set of
FF
\EQ
F_n^{\varphi}(\be_1,\be_2,\ldots,\be_n)\,=\,
<0| \varphi(0) |\be_1,\be_2, \ldots, \be_n>\,=\,
            \frac{1}{\sqrt{2}}\,\delta_{1,n}\,\,\,,
\label{ffphi1}
\EN
and its two-point euclidean correlator reduces to a Bessel function
\bea
<\varphi(R) \,\varphi(0)>_E \,&=&\,
 \sum_{n=0}^{\infty}\frac{1}{n!}
\int_{-\infty}^{+\infty} \frac{d\be_1}{2\pi}
\ldots
\int_{-\infty}^{+\infty} \frac{d\be_n}{2\pi}
\left| F_n(\be1,\ldots,\be_n)\right|^2
e^{-m R \sum_i \cosh \be_i}\nonumber\\
\,&= &
\frac{1}{\pi} K_0(mR)\,\,\, .
\eea
The absence of interaction implies that the composite operators
$\varphi^k /k!$ are simply defined by the following FF
\EQ
<0| \frac{\varphi^k(0)}{k!} |\be_1,\be_2, \ldots, \be_n>\,=\,
            \left(\frac{1}{\sqrt{2}}\right)^k\,\delta_{n,k}\,\,\,.
\label{compop}
\EN
The equation of motion is compatible with a class of stress-energy tensor
labelled by the free parameter $Q$ appearing in the definition of $\Theta(x)$
\EQ
\Theta(x)\,=\,2\pi\left(m^2 \varphi^2+\frac{Q}{\sqrt \pi} \Box \varphi\right)
\,\,\, .
\EN
In terms of FF we have
\bea
F_0^{\Theta} &\,=\,& 0 \,\,\, ,\nonumber \\
F_1^{\Theta} &\,=\,& -\sqrt{2 \pi} m^2 \,Q \,\,\, ,
\label{fftrace1}
\\
F_2^{\Theta} &\,=\,& 2 \pi m^2 \,\,\, ,\nonumber \\
F_k^{\Theta} &\,=\,& 0 \,\,\,\,\, , k>2 \,\,\,.\nonumber
\eea
The meaning of $Q$ becomes clear once we analyze the ultraviolet limit of
this massive theory. The central charge of the underlying CFT which rules
the ultraviolet properties of the model may be computed by using the
$c$-theorem sum rule \cite{cth,Cardycth}
\EQ
c\,=\,\frac{3}{2} \int_0^{\infty} dR \,R^3
             <\Theta(R)\Theta(0)>_E \,\,\, .
\label{cardy}
\EN
The euclidean correlator is given by
\bea
<\Theta(R) \Theta(0)>_E &\,=\,&
 \sum_{k=0}^{\infty}\frac{1}{k!}
\int_{-\infty}^{+\infty} \frac{d\be_1}{2\pi}
\ldots
\int_{-\infty}^{+\infty} \frac{d\be_k}{2\pi}
\left| F_k(\be1,\ldots,\be_k)\right|^2
e^{-m R \sum_i \cosh \be_i} \nonumber \\
&\,=\,&m^4 \left(2 \left(K_0(mR)\right)^2 \,+\,2  Q^2 K_0(mR)
\right)\,\,\, ,
\eea
and the result of the integral (\ref{cardy}) is simply
\EQ
c\,=\,1\,+ 12 \,Q ^2 \,\,\, .
\label{CC}
\EN
Hence the one-particle FF of $\Theta$ is related to the background charge
of the CFT reached in the ultraviolet limit. We could have obtained the same
result by directly analyzing the ultraviolet limit of the holomorphic
component of the stress-energy tensor. Indeed
\bea
<T_{z z}(z,\bar{z}) T_{z z}(0,0)>_E &=&
\left(\frac{\bar{z}}{z}\right)^2\,
<T_{z z}(R,R) T_{z z}(0,0)>_E \nonumber \\
&=&
m^4 \,\left(\frac{\bar{z}}{z}\right)^2\,
\left(2 \left( K_2(mR) \right)^2 + 2 Q^2 K_4(mR) \right)
\,\,\,,
\eea
($R^2=z\bar{z}$ and $z=x^0+i x^1$, $\overline z=x^0-i x^1$), and in the limit
$(mR) \goto 0$ we have
\EQ
<T_{z z}(z,\overline z) T_{zz}(0,0)>\,=\,
\frac{c}{2z^4}\,=\,\frac{1\,+ \,12 \,Q^2}{2z^4}\,\,\,.
\EN
To complete our discussion on the free theory, let us compute the conformal
dimensions $\Delta(\alpha)$ characterizing the scaling properties of the
exponential operators $V_a=e^{\alpha\varphi}$ in the ultraviolet limit.
This will be identified as the coefficient of the most singular term in
the ultraviolet limit of the correlator $<T_{zz}(z,\bar{z})\, V_{\alpha}(0)>$.
Using the FF of $V_{\alpha}(0)$ given by
\EQ
<0| V_{\alpha}(0)|\be_1,\ldots,\be_n>\,=\,
\left(\frac{\alpha}{\sqrt{2}}\right)^n\,\,\,,
\label{ffexp}
\EN
we have
\bea
<T_{z z}(z,\bar{z})\,V_{\alpha}(0)>_E &=&
\left(\frac{\bar{z}}{z}\right) \,
<T_{z z}(R, R)\,V_{\alpha}(0)>_E \nonumber \\
&=&
m^2\,\left(\frac{\bar{z}}{z}\right)\,
\left(- \frac{\alpha^2}{2 \pi} \left( K_1(mR) \right)^2
\,+\, \alpha \frac{Q}{\sqrt{\pi}} K_2(mR)  \right) \,\,\, ,
\nonumber
\eea
and for $(mR)\goto 0$
\EQ
<T_{z z}(z,\bar{z})\,V_{\alpha}(0)>_E \,\sim\,
\frac{1}{z^2} \left(- \frac{\alpha^2}{8\pi} \,+\,
\frac{\alpha\, Q}{\sqrt{4 \pi}} \right)
\,\,\, ,
\EN
i.e.
\EQ
\Delta\left(\alpha\right)\,=\,- \frac{\alpha^2}{8\pi} \,+\,
\frac{\alpha\,Q}{\sqrt{4 \pi}} \,\,\, .
\EN
Due to the background charge, they differ from the gaussian value
$\Delta(\alpha)=-\alpha^2/8\pi$.

\resection{The Sinh-Gordon Model}

The free theory provides an easy example of non-trivial ultraviolet behaviours
induced by a non-zero value of $F_1^{\Theta}$. It is interesting to see if
similar occurrences are also present for interacting theories. In this section
we will analyze the Sinh-Gordon model and in the next one the Bullough-Dodd
model.

\subsection{Basic Properties}

The Sinh-Gordon theory is a classical integrable model defined by the equation
of motion
\bea
\Box \varphi &\,=\,&-\, \frac{m_0^2}{2g}
\left( e^{g \varphi} \,- \,e^{-g \varphi}\right)\,\,\,.
\label{cem2}
\eea
The theory is invariant under a $Z_2$-symmetry $\varphi\goto -\varphi$. The
integrability of the model also persists at the quantum level and the
exact two-body elastic $S$-matrix involving the asymptotic particles created
by the operator $\varphi$ is given by \cite{AFZ}
\EQ
S(\beta,B)\,=\,
\frac{\tanh\frac{1}{2}(\beta-i\frac{\pi B}{2})}
{\tanh\frac{1}{2}(\beta+i\frac{\pi B}{2})} \label{smatrix}\,\,\,.
\EN
The coupling constant dependence of the model is encoded into the function
\EQ
B(g)\,=\,\frac{g^2}{4 \pi}\frac{1}{\left(1 + g^2/8 \pi\right)}\,\,\,.
\label{renormcoupling}
\EN
Since there is no pole on the physical sheet $0<{\rm Im}\,\beta<\pi$ for
real values of the coupling constant $g$, the Sinh-Gordon model presents no
bound states. Notice that the $S$-matrix is invariant under the duality
transformation $B\,\goto\,2-B$, which establishes a mapping of the theory
between the weak coupling and strong coupling regimes, i.e.
$g\rightarrow 8\pi/g$. This symmetry will be respected by all FF of manifestly
self-dual operators.

\subsection{Space of the Form Factors}

The form factors of the Sinh-Gordon model have been investigated in
\cite{FMS,KM}. In this subsection we briefly recall the basic results
obtained in refs.\,\cite{FMS,KM} which are relevant for our subsequent
considerations.

In order to compute the FF of this theory, the first step is to take into
account their monodromy properties dictated by the $S$-matrix. To this aim, let
$F_{\rm min}^{\rm SG}(\beta,B)$ be the solution of the functional equations
\EQ
\begin{array}{ccl}
F_{\rm min}^{\rm SG}(\beta,B)&=&F_{\rm min}^{\rm SG}(-\beta,B)\,
S(\beta,B)\,\, ,\\
F_{\rm min}^{\rm SG}(i\pi-\beta,B)&=&F_{\rm min}^{\rm SG}(i\pi+\beta,B)\,\, ,
\end{array}
\label{Watson2}
\EN
which has no poles and zeros in the physical sheet $0 <{\rm Im}\,\beta\leq\pi$
and with an asymptotic behaviour given by
\EQ
\lim_{\beta \rightarrow \infty} F_{\rm min}^{\rm SG}(\beta,B) = 1\,\,.
\label{limitefmin}
\EN
Its explicit expression reads
\begin{eqnarray}
F_{\rm min}^{\rm SG}(\beta,B) & = & {\cal N}(B)\,\exp\left[
8\int_0^{\infty} \frac{dx}{x} \frac{\sinh\left(\frac{x B}{4}\right)
\sinh\left(\frac{x}{2}(1-\frac{B}{2})\right) \,\sinh\frac{x}{2}}{\sinh^2 x}
\sin^2\left(\frac{x\hat\beta}{2\pi}\right)\right] \,\,\,, \nonumber\\
{\cal N}(B) & = & \exp\left[-4\int_0^{\infty}
\frac{dx}{x} \frac{\sinh\left(\frac{x B}{4}\right)
\sinh\left(\frac{x}{2}(1-\frac{B}{2})\right) \,\sinh\frac{x}{2}}{\sinh^2 x}
\right] \,\,\, ,
\label{integral}
\end{eqnarray}
where $\hat\beta=i\pi-\beta$. Equivalently
\EQ
F_{\rm min}^{\rm SG}(\beta,B)\,=\,
\prod_{k=0}^{\infty}
\left|
\frac{\Gamma\left(k+\frac{3}{2}+\frac{i\hat\beta}{2\pi}\right)
\Gamma\left(k+\frac{1}{2}+\frac{B}{4}+\frac{i\hat\beta}{2\pi}\right)
\Gamma\left(k+1-\frac{B}{4}+\frac{i\hat\beta}{2\pi}\right)}
{\Gamma\left(k+\frac{1}{2}+\frac{i\hat\beta}{2\pi}\right)
\Gamma\left(k+\frac{3}{2}-\frac{B}{4}+\frac{i\hat\beta}{2\pi}\right)
\Gamma\left(k+1+\frac{B}{4}+\frac{i\hat\beta}{2\pi}\right)}
\right|^2 \,\,\, .
\label{GammaSh}
\EN
Since $S(0,B)=-1$ (for $B\neq 0$ and $2$), $F_{\rm min}^{\rm SG}(\beta,B)$
vanishes at the two-particle threshold value $\beta=0$. As discussed in
\cite{Camus}, this generally induces a suppression of all higher thresholds
in the spectral representation of the correlation functions and
gives rise to very fast convergent series.

In terms of $F_{\rm min}^{\rm SG}(\beta,B)$, a convenient parameterization of
the $n$-particle FF of the Sinh-Gordon model is given by
\EQ
F_n(\beta_1,\ldots,\beta_n)\,=\, H_n\, Q_n(x_1,\ldots,x_n)\,
\prod_{i<j} \frac{F_{\rm min}^{\rm SG}(\beta_{ij},B)}{(x_i+x_j)}
\,\,\, ,\label{para}
\EN
Here $x_i\equiv e^{\beta_i}$ and $H_n$ are normalization constants, which
can be conveniently chosen as
\EQ
H_{2n+1} = H_{1}\mu^{2n}(B)\,\,\,, \,\,\, H_{2n}
= H_{2}\mu^{2n-2}(B)
\,\,\, ,
\label{normalconst}
\EN
with
\EQ
\mu(B) \equiv
\left ( \frac{4 \sin (\pi B /2)}{{\cal N}(B)}
\right )^{\frac 12} \,\,\, .
\label{mu}
\EN
The functions $Q_n(x_1,\dots,x_n)$ are symmetric polynomials in the variables
$x_1,\dots,x_n$, solutions of the recursion equations
\EQ
(-)^n\,Q_{n+2}(-x,x,x_1,\ldots,x_n)\, = \,x  D_n(x,x_1,x_2,\ldots ,x_n)
\,Q_n(x_1,x_2,\ldots,x_n) \,\,\, ,
\label{rec}
\EN
with
\EQ
D_n(x,x_1,\dots,x_n) = \sum_{k=1}^n \sum_{m=1,odd}^k [m]\, x^{2(n-k)+m}
\sigma_{k}^{(n)}\sigma_{k-m}^{(n)} (-1)^{k+1} \,\,\, ,
\label{D_n}
\EN
and
\EQ
[n]\equiv\frac{\sin (n\pi\frac B2)}{\sin(\pi\frac B2)} \,\,\, .
\EN
For FF of spinless operators, their total degree is equal to $n(n-1)/2$
whereas their partial degree in each variable $x_i$ is fixed by the asymptotic
behaviour of the operator ${\cal O}_k$ which is under investigation.

As shown in \cite{KM}, the problem to classify all possible scalar operators
of the Sinh-Gordon model reduces to find the most general class of solutions
of eq.\,(\ref{rec}). Since this is a homogeneous equation, its solutions span
a linear space whose basis may be written in terms of the so-called
{\em elementary solutions} given by
\EQ
{\cal Q}_{n}(k)\,=\, {\rm det}\, M_{ij}(k) \,\,\, ,
\EN
where $M_{ij}(k)$ is an $(n-1)\times (n-1)$ matrix with entries
\EQ
M_{ij}(k) =
\s_{2i-j}\, [i-j+k]\,\,\, .
\label{element}
\EN
These polynomials depend on an arbitrary integer $k$ and satisfy
\EQ
{\cal Q}_{n}(k)\,=\, (-1)^{n+1} {\cal Q}_{n}(-k) \label{pr1} \,\,\, .
\EN
Therefore the structure of the FF of the SGM consists in a sequence of finite
linear spaces whose dimensions grow linearly as $n$ increasing the number
$2n-1$ or $2n$ of external particles. The reason is that, at each level of
the recursive process, the space of the FF is enlarged by including the
kernel solutions of the recursive equation (\ref{rec}), i.e.
$Q_{n}(-x,x,x_1,\ldots,x_{n-2})\,=\,0$. With the constraint that the total
order of the polynomials is $\frac{n(n-1)}{2}$, the kernel is unique and
given by $\Sigma_n(x_1,\ldots,x_n)\,=\,{\rm det}\, \sigma_{2i-j}$. These
solutions then gives rise to the half-infinite chains under the recursive
pinching $x_1=-x_2=x$
\EQ
\begin{array}{ccccccccccccccc}
\ldots & \goto & Q_{n+4}^{(1)} & \goto & Q_{n+2}^{(1)} & \goto & Q_{n}^{(1)} &
\goto & Q_{n-2}^{(1)} &\goto & \ldots & \goto & Q_{3}^{(1)} &\goto & 1 \\
\ldots & \goto & Q_{n+4}^{(2)} & \goto & Q_{n+2}^{(2)} & \goto & Q_{n}^{(2)}
& \goto & Q_{n-2}^{(2)} &\goto & \ldots & \goto & \Sigma_2 & & \\
 & & & . & & .& & . & & . & & . &  & &  \\
 & & & . & & .& & . & & . & & . &  & &  \\
\ldots & \goto & Q_{n+4}^{(n-2)} & \goto & Q_{n+2}^{(n-2)} & \goto
& Q_{n}^{(n-2)} & \goto & \Sigma_{n-2} & &  & & & &  \\
\ldots & \goto & Q_{n+4}^{(n)} & \goto & Q_{n+2}^{(n)} & \goto
& \Sigma_n & & & &  & & & &  \\
\ldots & \goto & Q_{n+4}^{(n+2)} & \goto & \Sigma_{n+2} &
& & & & &  & & & &
\end{array}
\label{chain}
\EN
The explicit expressions of such solutions can be found by determining
the linear combination of ${\cal Q}_n(k)$ which reduces to $\Sigma_n$
at the level $n$.

\subsection{Cluster Operators and Fundamental Exponentials}

The {\em fundamental exponential operators}
$\Phi_{\pm}(x)=e^{\pm g\varphi(x)}$ define the Sinh-Gordon model and
in general appear in the expression of the stress-energy tensor. In order
to calculate their matrix elements, let us consider initially those FF which
satisfy the requirements
\begin{itemize}
\item
To be asymptotically constant for $\beta_i\rightarrow \infty$, i.e.
\[
F_n(\be_1 + \Delta,\be_2,\ldots,\be_n)\,\,
\stackrel{\Delta \goto \infty}{\longrightarrow}\,\,o(1)
\,\,\,.
\]
\item To be proportional to the combination\footnote{As discussed in sect.\,2,
this factorization property is shared by the general FF of $\Theta$.}
$\sigma_1\sigma_{n-1}$ (for $n>2$).
\item
To be the solution of the {\em cluster equations}
\[
\lim_{\Delta \goto + \infty} F_{k+l} \left(\be_1+\Delta,\ldots,
     \be_k+\Delta,\be_{k+1},\ldots,\be_{k+l}\right)\,=\,
F_k \left(\be_1,\ldots,\be_k \right)
F_l \left(\be_{k+1},\ldots,\be_{k+l} \right)
\,\,\,
\]
with initial condition $F_0\,=\,1$.
\end{itemize}
There are two classes of FF which fulfill the three above conditions. Their
expressions are given by
\EQ
F_n^{(\pm)}(\beta_1,\ldots,\beta_n)\,=\,H_n^{(\pm)}(B)
\,Q_n(1) \,\prod_{i<j}^n\frac{F_{\rm min}^{\rm SG}(\beta_{ij},B)}{(x_i+x_j)}
\,\,\, ,
\label{twodime}
\EN
where
\EQ
H_n^{(+)}(B) \,=\,(\mu(B))^n
\,\,\,\,\, ,\,\,\,\,\, H_n^{(-)}\,=\,(-1)^n\,(\mu(B))^n \,\,\, .
\label{h12}
\EN
The corresponding operators, which are self-dual by construction, will be
called {\em cluster operators} and denoted as $V_{\pm}(x,B)$. We conjecture
that the fundamental exponentials of the Sinh-Gordon model may
be written as\footnote{For the value of the step function at the origin we
use $\theta(0)=1/2$.}
\EQ
\begin{array}{l}
\Phi_{+}(x,B)\,\equiv\, \theta(1-B) \,V_+(x,B) \,+\, \theta(B-1) \,V_-(x,B)
\,\,\, ,\\
\Phi_{-}(x,B)\,\equiv\, \theta(1-B) \,V_-(x,B) \,+ \,\theta(B-1) \,V_+(x,B)
\,\,\,.
\end{array}
\label{fundamental}
\EN
Postponing a non-trivial check of this position until when we will study the
UV-behaviour of the model, let us in the meantime discuss the properties of
the operators $\Phi_{\pm}(x,B)$ so defined.

First of all, they satisfy the cluster property by construction, in agreement
with the perturbative analysis for the matrix elements of the operators
$e^{\pm g\varphi(x)}$. Secondly, the FF of $\Phi_{\pm}(x,B)$ are not
individually invariant under the duality transformation but each
operator is mapped onto the other under the mapping $B\,\goto\,2-B$, i.e.
\EQ
\Phi_{\pm}(x,B)\,=\,\Phi_{\mp}(x,2-B)
\,\,\,.
\EN
Therefore they form a bidimensional representation of the duality symmetry.
However, this mapping becomes degenerate at the self-dual point $B=1$
where an identification occurs between the two exponential operators
$\Phi_{\pm}(x,B)$. Namely, at $B=1$ the matrix elements of the two fundamental
exponentials become indistinct and denoting by $\Phi(x)$ the resulting
operator, its FF are given by
\EQ
F_{n}^{\Phi}(\beta_1,\ldots,\beta_{2})\,=\,
\left\{
\begin{array}{ll}
\left(\mu(1)\right)^{n}
\,Q_{n}(1) \,\prod_{i<j}^n F_{\rm min}^{\rm SG}(\beta_{ij})/(x_i+x_j)
\,&
\mbox{$n$ even} \,\,\, ,\\
0 & \mbox{$n$ odd} \,\,\, .
\end{array}
\right.
\label{FFSD}
\EN
The identification of $\Phi_{\pm}(x)$ at the self-dual point has the
additional consequence that the resulting field $\Phi(x)$ is an even
operator under the $Z_2$ parity of the Sinh-Gordon model, as it is evident
from the vanishing of its matrix elements on all $2n+1$ particle states.

Using the FF of the fundamental exponentials and those of the elementary field
given by
\EQ
F_n^{\varphi}(\beta_1,\ldots,\beta_n)\,=\,
H_{n}^{\varphi}\,Q_n(0) \,\prod_{i<j}^n\frac{F_{\rm min}^{\rm SG}(\beta_{ij})}
{x_i+x_j}
\EN
\[
H_{2n+1}^{\varphi}\,=\,\frac{1}{\sqrt 2} \left(\mu(B)\right)^{n} \hspace{3mm} ,
\hspace{3mm} H_{2n}^{\varphi}=0 \,\,\, ,
\]
the quantum version of the equation of motion may be written as
\bea
\Box \varphi(x) &\,=\,&\frac{m^2}{2\sqrt{2}\mu(B)} \,
 \left(\theta(1-B)\,-\,\theta(B-1) \right)
 \left(e^{-g\varphi(x)} - e^{g\varphi(x)}\right)
\nonumber \\
&\,=\,&\frac{m^2}{2\sqrt{2}\mu(B)} \,
 \left(V_-(x,B) - V_+(x,B) \right)\,\,\,.
\label{sgqem}
\eea
This equation has to be understood as an identity satisfied by the FF of
the operators appearing on the left and right sides of this relation.

\subsection{Class of Stress-Energy Tensors}

The quantum equation of motion (\ref{sgqem}) is compatible with a
one-dimensional space of stress-energy tensors given by
\EQ
\Theta(x) \,=\, F_0^{\Theta}(B)\,
\left(a\,\Phi_+(x,B) + (1\,-\,a)\,\Phi_-(x,B)\right)
\,\,\,.
\label{tetasg}
\EN
The normalization constant $F_0^{\Theta}(B)$ may be fixed by means of the
Thermodynamical Bethe Ansatz \cite{TBA}
\EQ
F_0^{\Theta}(B) \,=\, \frac{\pi m^2}{2\sin(\pi B/2)} \,\,\,.
\EN
The variable $a$, on the contrary, is a free parameter. Varying its value,
we can weight differently the two fundamental exponentials in the trace and,
consequently, we can interpolate between different scaling regimes of the
Sinh-Gordon model in its ultraviolet limit.

\subsubsection{The case $a=1$}

The trace of the stress-energy tensor is given in this case by
\EQ
\Theta(x)\,=\,F_0^{\Theta}(B)\,\Phi_{+}(x,B)\,\,\,.
\label{pos1}
\EN
With such definition of $\Theta$, we expect that the massive theory will
flow in the ultraviolet regime to a CFT defined by the bare action
\EQ
{\cal S}_-\,=\,\int d^2x \left[
\frac{1}{2}(\partial_{\mu}\varphi)^2-\frac{m_0^2}{2 g^2}
\,e^{- g \varphi}\,\right]\,\,.
\label{action1}
\EN
In order to support this conclusion, let us compare the central charge
associated to the CFT (\ref{action1}) with the central charge obtained,
on the contrary, in terms of the FF by using the $c$-theorem sum rule
(\ref{cardy}).

Assuming eq.\,(\ref{action1}) as definition of the ultraviolet theory,
the corresponding central charge is given by (see, for instance \cite{HM})
\EQ
c(g)\,=\,1 \,+\, 12 \,Q_-^2(g)
\EN
where
\EQ
Q_-(g)\,=\,-\,\left(\frac{\sqrt{4\pi}}{g} \,+\, \frac{g}{2\sqrt{4 \pi}}
\right) \,\,\, .
\label{chargeinfi1}
\EN
Using eq.\,(\ref{renormcoupling}), it may be written as
\EQ
c(B) \,=\, 1\,+\, 6 \left( \frac{2-B}{B}
\,+\, \frac{B}{2-B}\,+\,2 \right)\,\,\,.
\label{guessc1}
\EN
Notice that this is a self-dual function of the coupling constant, i.e.
invariant under $B\goto 2-B$.

On the other hand, we may compute the central charge associated to the
ultraviolet limit of the massive theory in terms of the second moment of the
two-point function of the trace $\Theta(x)$ \cite{cth,Cardycth}
\EQ
c(B)\,=\,\frac{3}{2} \int_0^{\infty} dR \,R^3
             <\Theta(R)\Theta(0)>_E \,\,\,.
\label{cardy2}
\EN
According to eq.\,(\ref{pos1}), the two-point function of the trace $\Theta(x)$
has to be computed in terms of the FF of the operator $\Phi_{+}(x,B)$
defined in eq.\,(\ref{fundamental}). The data reported in Table 1 and plotted
in Fig.\,1 show that the first two FF of $\Phi_+(x)$ are sufficient to
saturate the sum-rule (\ref{cardy2}) and to reproduce with high percentage of
precision the expression (\ref{guessc1}).

An additional check that the ultraviolet limit induced by this choice of
$\Theta$ is ruled by CFT (\ref{action1}), is given by the computation
of the conformal dimensions of the fundamental exponentials. This can be
done in two different ways, using directly CFT method or analyzing the
ultraviolet behaviour of massive correlators.

For the CFT defined by eq.\,(\ref{action1}), the conformal dimensions of the
primary fields corresponding to the operators $e^{\alpha \varphi}$ are given by
\EQ
\Delta_-(\alpha)\,=\, - \frac{\alpha^2}{8 \pi} \,+\,
\frac{\alpha Q_-(g)}{\sqrt{4 \pi}} \,\,\, ,
\label{andim(-g)2}
\EN
and therefore, for the fundamental exponentials we have
\EQ
\begin{array}{ll}
\Delta_-(\Phi_-) &\,=\, 1 \,\,\, ,\\
\Delta_-(\Phi_+) &\,=\, -\,1\,-\,\frac{g^2}{4 \pi} \,\,\,.
\end{array}
\label{anom1}
\EN
On the other hand, we may compute the conformal dimensions of the fundamental
exponentials by investigating the UV-limit of the correlators
\bea
<T_{z z}(z, \bar{z})\,\Phi_{\pm}(0)>_E &=&
\frac{\bar{z}}{z}\,
 \sum_{n=0}^{\infty}\frac{1}{n!}
\int_{-\infty}^{+\infty} \frac{d\be_1}{2\pi}
\ldots
\int_{-\infty}^{+\infty} \frac{d\be_n}{2\pi}
\left( F_n^{T_{+ +}}(\be1,\ldots,\be_n)\right)^{*}
\nonumber \\
& & \,\,\,\,\,\,\,\,\,\,\,\,\,\,
\times \, F_n^{\Phi_{\pm}}(\be1,\ldots,\be_n)
e^{-m R \sum_i \cosh \be_i}
\,\,\,.
\eea
At order $O(g^4)$, it is sufficient to truncate the series to the first two
terms and also use the perturbative expansion
\EQ
{\cal N}(B)= 1 - \frac{g^2}{8 \pi^2} + O(g^4)\,\,\,.
\EN
Since for small values of $g$
\EQ
F_n^{T_{+ +}}(x_1,\ldots,x_n) \,=\, -F_0^{\Theta}(B)\,
\frac{\sigma_{n-1}^{(n)}}{4\sigma_1^{(n)}\sigma_n^{(n)}}
\,F_n^{V_+}(x_1,\ldots,x_n)\,\,\, ,
\EN
and
\bea
<T_{z z}(z,\bar{z})\,\Phi_{\pm}(0)>_E &\simeq&
\frac{\bar{z}}{z} \left\{ -\,F_0^{\Theta}(B)
\,F_1^{V_+} F_1^{V_{\pm}} \frac{1}{4 \pi} K_2(mR) \right.
\\
& & \left.
\,-\, \frac{F_0^{\Theta}(B)}{2 \pi^2} \int_{-\infty}^{+\infty} d\be\,
K_2(2mR \cosh \be)\,F_2^{V_+}(2\be) F_2^{V_{\pm}}(2\be)
\right\} \,\,\, , \nonumber
\eea
in the limit $mR \goto 0$, we have
\EQ
<T_{z z}(z,\bar{z})\,\Phi_{\pm}(0)>_E\,\,
\simeq \,\frac{\Delta_-(\Phi_{\pm})}{z^2}
\,\,\, ,
\EN
with
\EQ
\begin{array}{ll}
\Delta_-(\Phi_-) &\,=\, 1 +O(g^4) \,\,\, ,\\
\Delta_-(\Phi_+) &\,=\, -\,1\,-\,\frac{g^2}{4\pi} + O(g^4)\,\,\, ,
\end{array}
\EN
in agreement with eq.\,(\ref{anom1}).

\subsubsection{The case $a=0$}

The trace of stress-energy tensor is given in this case by
\EQ
\Theta(x)\,=\, F_0^{\Theta}(B)\,\Phi_{-}(x,B)\,\,\, .
\label{pos2}
\EN
and we expect that the massive model will flow in the ultraviolet limit to
a CFT defined by the bare action
\EQ
{\cal S}_+\,=\,\int d^2x \left[
\frac{1}{2}(\partial_{\mu}\varphi)^2-\frac{m^2}{2 g^2}
\,e^{g \varphi}\,\right]\,\,\, .
\label{action2}
\EN
The corresponding background charge is given by \cite{HM}
\EQ
Q_+(g)\,=\,\left(\frac{\sqrt{4\pi}}{g} + \frac{g}{2\sqrt{4 \pi}}\right)\,\,\, .
\EN
This CFT differs from that analyzed in the previous subsection for the
exchange of the role of the two fundamental exponentials.

According to the CFT defined by the bare lagrangian (\ref{action2}), the
anomalous dimensions of the primary fields corresponding to the operators
$e^{\alpha\phi}$ are given
by
\EQ
\Delta_+(\alpha)\,=\, - \frac{\alpha^2}{8 \pi} \,+\,
\frac{\alpha Q_+(g)}{\sqrt{4 \pi}} \,\,\, ,
\label{andim(+g)2}
\EN
and for the fundamental exponentials we have in this case
\EQ
\begin{array}{ll}
\Delta_+(\Phi_+) &\,=\, 1  \,\,, \\
\Delta_+(\Phi_-) &\,=\, -\,1\,-\,\frac{g^2}{4\pi} \,\, .
\end{array}
\EN
Repeating the same kind of computations of the previous subsection, it is easy
to check that these expressions coincide, at order $O(g^4)$, with the
conformal dimensions extracted from the ultraviolet behaviour of the
correlators $<T_{zz}(z,\overline z)\,\Phi_{\pm}(0)>$.

For what concerns the central charge, since it depends quadratically on
$Q_{\pm}$, its value is given as before by the self-dual function
(\ref{guessc1}). Analogous computations for the second moment of the
$\Theta$ computed in terms of the FF of $\Phi_{-}(x,B)$ (which is the dual
operator of $\Phi_+(x,B)$), lead therefore to the same results of Table 1
(see also Fig.\,1).

\subsubsection{General case}

We are now able to give the conformal dimension $\Delta(\alpha)$ of the
exponential operator $e^{\alpha \varphi}$ and the central charge of the CFT
reached in the ultraviolet regime for generic value of the parameter $a$.

Since $\Delta(\alpha)$ is the coefficient of the most singular term obtained
in the UV-limit of the correlation function
$<T_{z z}(z,\overline z)\, e^{\alpha\varphi(0)}>$, and the FF of
$T_{zz}(z,\overline z)$ depends linearly on the parameter $a$,
the conformal dimension is given by
\begin{eqnarray}
\Delta(\alpha) & =&\, a\,\Delta_-(\alpha)\,+\,(1-a)\,\Delta_+(\alpha)\,=
\label{andimsg}\\
& = & - \frac{\alpha^2}{8 \pi} \,+\,
\frac{\alpha}{\sqrt{4 \pi}}
\,(1\,-\,2 a) \left(\frac{\sqrt{4 \pi}}{g} \,+\,
\frac{g}{2\sqrt{4\pi}} \right)
\,\,\,\,,\nonumber
\end{eqnarray}
with $\Delta_{\pm}(\alpha)$ defined in eqs.\,(\ref{andim(-g)2}) and
(\ref{andim(+g)2}). The coefficient in front of the linear term in $\alpha$
in eq.\,(\ref{andimsg}) identifies the background charge and therefore
the central charge of the CFT reached in the ultraviolet limit is given by
\EQ
c\,=\,1\,+\,24\,\frac{(1-2a)^2}{B(2-B)}
\,\,\,.
\label{gencuv}
\EN
We have checked the validity of this result with the computation of the
central charge in terms of the first FF of the operator (\ref{tetasg}).
The comparison between them is shown in Fig.\,3, varying $a$ at fixed $B$.

As last example of possible choices of the stress-energy tensor, observe
that for $a=1/2$ we have $c=1$, independent of the coupling constant. The
corresponding expression of $\Theta$ is given by \cite{FMS}
\EQ
\Theta(x) \,=\, \frac{F_0^{\Theta}(B)}{2}
\left( e^{g \varphi} \,+\, e^{- g \varphi}\right)
\,\,\,.
\EN
This operator is manifestly self-dual and $Z_2$-even. With this choice
of $a$, the anomalous dimensions of the fundamental exponentials coincide,
at lowest order, with their gaussian values
\EQ
\Delta(\pm g) \,=\, -\,\frac{g^2}{8 \pi} \,+\,o(g^4)\,\,\,.
\EN
The check of the central charge obtained in this case has been done in the
two-particle approximation in ref.\,\cite{FMS} and is reported here in table 2.

\resection{The Bullough-Dodd massive boson}

Different ultraviolet scaling regimes induced by different choice of the
stress-energy tensor can be easily discussed for another integrable theory
involving an interacting bosonic field, the so-called Bullough-Dodd (BD) model.

\subsection{Basic Properties}

The Bullough-Dodd (BD) model is defined by the equation of motion
\cite{DB,ZS,MOP}
\bea
\Box \varphi &\,=\,& \frac{m_0^2}{3\lambda}
\left( e^{- 2 \lambda \varphi} - e^{\lambda \varphi}\right)\,\,\,.
\label{cem3}
\eea
At the quantum level, the integrability of the model leads to the elasticity
and factorization of the scattering processes. The spectrum of the model
consists of a massive particle state $A$ created by the elementary field
$\varphi$. This particle appears as bound state of itself in the scattering
process
\EQ
A\,\times\, A\,\rightarrow A\,\rightarrow A\,\times\, A \,\,\,.
\label{boot}
\EN
The corresponding S-matrix is given by \cite{AFZ}
\EQ
S(\beta,{\cal B})\,=\,
f_{\frac{2}{3}}(\beta)\,f_{\frac{{\cal B}}{3}-\frac{2}{3}}(\beta)\,
f_{-\frac{{\cal B}}{3}}(\beta)\,\,\,.
\label{BDD}
\EN
Here
\EQ
f_x(\beta)\,=\,\frac{\tanh\frac{1}{2}(\beta+i\pi x)}
{\tanh\frac{1}{2}(\beta-i\pi x)}\,\,\,  ,
\EN
and the coupling constant dependence of the model is encoded into the function
\EQ
{\cal B}(\lambda)\,=\,
\frac{\lambda^2}{2\pi}\frac{1}{1+\frac{\lambda^2}{4\pi}}\,\,\,.
\label{calB}
\EN
Like the Sinh-Gordon model, the $S$-matrix of the BD model is invariant
under the mapping
\EQ
{\cal B}(\lambda)\,\rightarrow\, 2-{\cal B}(\lambda) \,\,\, ,
\EN
i.e. under the weak-strong coupling constant duality
$\lambda \, \rightarrow \,4\pi/\lambda$.

The minimal part of the $S$-matrix, i.e. the term $f_{\frac{2}{3}}(\beta)$,
contains the physical pole $\beta=2\pi i/3$ of the bound state and, as matter
of fact, it  coincides with the $S$-matrix of the Yang-Lee model
\cite{Cardymus}. Taking into account the coupling constant dependence of the
$S$-matrix, the residue at the pole is given by
\EQ
\Gamma^2({\cal B})\,=\,2{\sqrt 3} \frac{\tan\left(\frac{\pi {\cal
B}}{6}\right)}
{\tan\left(\frac{\pi {\cal B}}{6}-\frac{2\pi}{3}\right)}
\frac{\tan\left(\frac{\pi}{3}-\frac{\pi {\cal B}}{6}\right)}
{\tan\left(\frac{\pi {\cal B}}{6}+\frac{\pi}{3}\right)}\,\,\,.
\label{eq: threepv}
\EN
This function, that corresponds to the three-particle vertex on mass-shell,
vanishes for ${\cal B}=0$ and ${\cal B}=2$ (which are the free theory limits)
with the corresponding scattering amplitude $S=1$. However, it also vanishes
at the self-dual point ${\cal B}=1$, with the corresponding scattering
amplitude $S(\beta)=f_{-2/3}$. This coincides with the $S$-matrix
of the Sinh-Gordon model computed at $B=2/3$. As analyzed in \cite{FMS2}, this
equality between the $S$-matrices of the two models implies that at the
self-dual point the BD model dynamically develops a $Z_2$-symmetry which
is a non-perturbative property of the model.

\subsection{Form Factors}

Taking into account the bound state pole in the two-particle channel at
$\beta_{ij}=2\pi i/3$ and the one-particle pole in the three-particle channel
at $\beta_{ij}=i\pi$, the general form factors of the BD model can be
parameterized as
\EQ
F^k_n(\beta_1,\ldots,\beta_n)\,=
\, Q_n(x_1,\ldots,x_n)\,
\prod_{i<j} \frac{F_{\rm min}^{\rm BD}(\beta_{ij})}{(x_i+x_j)(\omega x_i+x_j)
(\omega^{-1}x_i+x_j)} \,\, , \label{para1}
\EN
where we have introduced the variables
\EQ
x_i\,=\,e^{\beta_i} \,\, , \,\, \omega\,=\,e^{i\pi/3}\,\, .
\EN
$F_{\rm min}^{\rm BD}(\beta)$ is an analytic function without zeros
and poles in the physical sheet, whose explicit expression is given by
\begin{eqnarray}
&& F_{\rm min}^{\rm BD}(\beta,{\cal B})\,=\,
\prod_{k=0}^{\infty}
\left|
\frac{\Gamma\left(k+\frac{3}{2}+\frac{i\hat\beta}{2\pi}\right)
\Gamma\left(k+\frac{7}{6}+\frac{i\hat\beta}{2\pi}\right)
\Gamma\left(k+\frac{4}{3}+\frac{i\hat\beta}{2\pi}\right)}
{\Gamma\left(k+\frac{1}{2}+\frac{i\hat\beta}{2\pi}\right)
\Gamma\left(k+\frac{5}{6}+\frac{i\hat\beta}{2\pi}\right)
\Gamma\left(k+\frac{2}{3}+\frac{i\hat\beta}{2\pi}\right)}
\right.
\\
&&\,\,\,\times\,\,\left.
\frac{\Gamma\left(k+\frac{5}{6}-\frac{B}{6}+\frac{i\hat\beta}{2\pi}\right)
\Gamma\left(k+\frac{1}{2}+\frac{B}{6}+\frac{i\hat\beta}{2\pi}\right)
\Gamma\left(k+1-\frac{B}{6}+\frac{i\hat\beta}{2\pi}\right)
\Gamma\left(k+\frac{2}{3}+\frac{B}{6}+\frac{i\hat\beta}{2\pi}\right)}
{\Gamma\left(k+\frac{7}{6}+\frac{B}{6}+\frac{i\hat\beta}{2\pi}\right)
\Gamma\left(k+\frac{3}{2}-\frac{B}{6}+\frac{i\hat\beta}{2\pi}\right)
\Gamma\left(k+1+\frac{B}{6}+\frac{i\hat\beta}{2\pi}\right)
\Gamma\left(k+\frac{4}{3}-\frac{B}{6}+\frac{i\hat\beta}{2\pi}\right)}
\nonumber \right|^2
\end{eqnarray}
($\hat\beta=i\pi-\beta$). Its normalization is fixed by requiring
the asymptotic behaviour
\EQ
\lim_{\beta \rightarrow \infty} F_{\rm min}^{\rm BD}(\beta,{\cal B}) = 1\,\,\,
{}.
\EN
Notice that at the self-dual point ${\cal B}=1$, the above function coincides
with the $F_{\rm min}^{\rm SG}\left(\beta,\frac{2}{3}\right)$ of the
Sinh-Gordon model, i.e.
\EQ
F_{\rm min}^{\rm BD}(\beta,1)\,=\,F_{\rm min}^{\rm SG}
\left(\beta,\frac{2}{3}\right) \,\,\,.
\label{BDvsSG}
\EN
The functions $Q_n(x_1,\dots,x_n)$ are symmetric polynomials in the variables
$x_1,\dots,x_n$. Using the functional relations satisfied by
$F_{\rm min}(\beta,{\cal B})$
\begin{eqnarray}
F_{\rm min}^{\rm BD}(i\pi+\beta,{\cal B})
\,F_{\rm min}^{\rm BD}(\beta,{\cal B})\,&=&\,
\frac{\sinh\beta\left(\sinh\beta+\sinh\frac{i\pi}{3}\right)}
{\left(\sinh\beta+\sinh\frac{i\pi {\cal B}}{3}\right)
\left(\sinh\beta+\sinh\frac{i\pi(1+{\cal B})}{3}\right)} \,\,\, ,
\nonumber\\
F_{\rm min}^{\rm BD}(\beta+\frac{i\pi}{3},{\cal B}) \,
F_{\rm min}^{\rm BD}(\beta-\frac{i\pi}{3},{\cal B})\,&=&\,
\frac{\cosh\beta+\cosh\frac{2i\pi}{3}}
{\cosh\beta+\cosh\frac{i\pi(2+{\cal B})}{3}}
\,\,F_{\rm min}^{\rm BD}(\beta,{\cal B}) \label{shift}\,\,\, ,
\end{eqnarray}
the kinematical and bound state residue conditions give rise to the following
recursive equations satisfied by the functions $Q_n(x_1,\dots,x_n)$ \cite{FMS2}
\EQ
(-1)^{n} Q_{n+2}(-x,x,x_1,x_2,\ldots,x_n)\,=
\,\frac{1}{F_{\rm min}^{\rm BD}(i\pi,{\cal B})}\,x^3 \,U(x,x_1,x_2,\ldots,x_n)
Q_n(x_1,x_2,\ldots,x_n)\,\,\, ,
\label{recc1}
\EN
where
\bea
U(x,x_1,\ldots,x_n) &=& 2\sum_{k_1,\ldots,k_6=0}^n
(-1)^{k_2+k_3+k_5}\, x^{6n-(k_1+\cdots\, +k_6)}
 \,\sigma_{k_1}^{(n)}\sigma_{k_2}^{(n)}\ldots\sigma_{k_6}^{(n)} \, \\
&&\,\,\, \times\,
\sin\left[\frac{\pi}{3}\left[2(k_2+k_4-k_1-k_3)+{\cal
B}(k_3+k_6-k_4-k_5)\right]
\right] \,\,\, ,\nonumber
\eea
and
\EQ
Q_{n+2}(\omega x,\omega^{-1}x,x_1,\ldots ,x_n)\,=\,-\,
\frac{\sqrt{3}}{F_{\rm min}^{\rm BD}\left(\frac{2\pi i}{3},{\cal B}\right)} \,
\Gamma({\cal B}) \,x^3 D(x,x_1,\ldots,x_n) \,Q_{n+1}(x,x_1,\ldots,x_n) \,\,\, ,
\label{recc2}
\EN
where
\bea
D(x,x_1,\ldots,x_n)\,&=&\,
\prod_{i=1}^n (x+x_i)(x \omega^{2+{\cal B}} + x_i)
(x \omega^{-{\cal B}-2} + x_i ) \\
 &=&\,
\sum_{k_1,k_2,k_6=0}^n
x^{3n-(k_1+k_2+k_6)} \, \omega^{(2+{\cal B})(k_2-k_3)} \,
 \,\sigma_{k_1}^{(n)}\sigma_{k_2}^{(n)}\sigma_{k_3}^{(n)} \,\,\,.
\nonumber
\eea

\subsection{Cluster Operators and Fundamental Exponentials}

Unlike the Sinh-Gordon model, we do not know presently a close solution for the
recursive equations satisfied by the FF of the BD model at generic value of
the coupling constant. However, as it will become clear in the following, all
we need for our consideration is the explicit computation of the first
representative FF of the elementary field $\varphi$ and of the so-called
cluster operators ${\cal V}_{\pm}(x)$.

The elementary field $\varphi$ is identified as that operator that creates
one-particle state. Therefore its lowest matrix element is given by
\EQ
F_1^{\varphi}\,=\,<0|\varphi(0)|\beta>\,=\,\frac{1}{\sqrt 2}\,\,\,.
\EN
With such normalization, the next FF are given by
\EQ
F_2^{\varphi}(\be)\,=\,-\,\frac{\Gamma({\cal B})}{\sqrt{2}}\,
\frac{\sin\frac{\pi}{6}(2+{\cal B})}{\sqrt{F_{\rm min}^{\rm BD}
(i\pi,{\cal B})}}\,
\frac{F_{\rm min}^{\rm BD}(\be,{\cal B})}{\cosh \be \,+ \frac{1}{2}}
\,\,\,,
\EN
and
\bea
F_3^{\varphi}(\beta_1,\beta_2,\beta_3)&=&
\left( \prod_{i<j}^3 \frac{F_{\rm min}^{\rm BD}(\be_i-\be_j,{\cal B})}
{(x_i+x_j)
(e^{i\pi/3} x_i+x_j)
(e^{-i\pi/3} x_i+x_j)}\right)
\, \frac{2\sqrt{2}}{F_{\rm min}^{\rm BD}(i\pi,{\cal B})}
\s_3\,\times  \nonumber \\
 & & \,\,\,\,
\left\{2 \sin^2 \frac{\pi}{6}(2+{\cal B})\,\Gamma^2({\cal B})\,
\left(\cos\frac{\pi}{3}(2+{\cal B})\,-1 \right)\,\s_3 \s_2 \s_1\,-\,
\right. \\
& & \,\,\left.
W({\cal B})\,\left(\s_3 \s_1^3 +\s_2^3 \right)\,+\,
\left(\sin^2 \frac{\pi}{6}(2+{\cal B})\,\Gamma^2({\cal B})\,+W({\cal B})
\right)
\s_2^2\,\s_1^2
\right\}\nonumber
\,\, ,
\eea
where
\EQ
W({\cal B})\,=\,2 \sqrt{3} \, \sin\left(\frac{\pi {\cal B}}{6} \right)\,
\sin\left(\frac{\pi }{6} (2-{\cal B}) \right)\,\,\,.
\EN
In terms of them, we can easily obtain the first FF of the operator
$\Box \varphi$
\EQ
F_n^{\Box \varphi}\,=\,-\,m^2\,\frac{\s_1 \s_{n-1}}{\s_n}
\,F_n^{\varphi}
\,\,\,.
\EN

In order to define the FF of the two fundamental exponential operators
$\Phi_1(x,{\cal B})\equiv e^{\lambda\varphi(x)}$ and
$\Phi_2(x,{\cal B})\equiv e^{-2\lambda\varphi(x)}$
of the Bullough-Dodd model, let us consider initially the definition of the
cluster operators ${\cal V}_{\pm}(x,{\cal B})$. As for the Sinh-Gordon model,
we are looking for these operators in the class of FF which are asymptotically
constant for $x_i\rightarrow\infty$ and proportional to the invariant
combination of symmetric polynomials $\s_1\sigma_{n-1}$ for $n>2$. The first
representative of such FF are given in Appendix A. The important point is that
all the higher FF obtained by solving the recursive equations will depend on
the constants $H_1$ and $H_2$ appearing in the equations (\ref{ff2class3})
and (\ref{ff3class3}), which play the role of arbitrary initial conditions of
the recursive structure. Their relative value can be fixed though, if we
require that the above FF satisfy an additional condition, i.e. the cluster
property
\EQ
\lim_{\Delta \goto + \infty} F_{k+l} \left(\be_1+\Delta,\ldots,
     \be_k+\Delta,\be_{k+1},\ldots,\be_{k+l}\right)\,=\,
F_k \left(\be_1,\ldots,\be_k \right)
F_l \left(\be_{k+1},\ldots,\be_{k+l} \right)
\,\,\,
\label{cccluster}
\EN
with $F_0\,=\,1$. In this case we have
\begin{eqnarray}
H_1^{\pm}({\cal B}) & = &\frac{1}{\sqrt{F_{\rm min}^{\rm BD}(i\pi,{\cal B})}}
\left\{
- \sin \left(\frac{\pi}{6}({\cal B}+2)\right) \Gamma({\cal B}) \pm
\sqrt{ \sin^2 \left(\frac{\pi}{6}({\cal B}+2)\right) \Gamma^2({\cal B})
       + 4 W({\cal B})} \right\} \,\,\, , \nonumber \\
H_2^{\pm}({\cal B}) & = & (H_1^{\pm})^2({\cal B}) \,\,\, .\label{branch}
\end{eqnarray}
With this choice of $H_1^{\pm}({\cal B})$ and $H_2^{\pm}({\cal B})$, the
infinite tower of FF with (\ref{ff2class3}) and (\ref{ff3class3})
as first representatives, define two cluster operators
${\cal V}_{\pm}(x,{\cal B})$. By construction, the matrix elements of such
operators are invariant under the duality transformation
${\cal B}\goto 2-{\cal B}$. Also in this case, we conjecture that the
fundamental exponential operators are given by
\EQ
\begin{array}{l}
\Phi_1(x,{\cal B})\,=\,\equiv\,
\theta(1-B) \,{\cal V}_+(x,{\cal B}) \,+\, \theta(B-1) \,{\cal V}_-(x,{\cal B})
\,\,\, ,\\
\Phi_2(x,{\cal B})\,=\,\equiv\,
\theta(1-B) \,{\cal V}_-(x,{\cal B}) \,+ \,\theta(B-1) \,{\cal V}_+(x,{\cal B})
\,\,\,.
\end{array}
\label{fundamentalBD}
\EN
This definition is in agreement with the perturbative analysis of the matrix
elements of the two exponential operators $e^{\lambda\varphi(x)}$ and
$e^{-2\lambda\varphi(x)}$. Concerning their properties under the duality
mapping, as far as ${\cal B}\neq 1$, the fundamental exponentials are mapped
each into the other under the mapping ${\cal B}\goto 2-{\cal B}$, i.e.
\EQ
\Phi_{1,2}(x,{\cal B})\,=\,\Phi_{2,1}(x,2-{\cal B})
\,\,\,.
\EN
However, this mapping becomes degenerate at the self-dual point ${\cal B}=1$
where, similarly to the Sinh-Gordon model, the two operators
$\Phi_1(x,{\cal B})$ and $\Phi_2(x,{\cal B})$ collapse into a single operator
$\Phi(x)$. Moreover, as already noticed in \cite{FMS2}, at the self-dual point
a $Z_2$-symmetry is dynamically implemented in the BD model. Due to the fact
that at ${\cal B}=1$ the three-particle vertex $\Gamma({\cal B})$ vanishes and
$H_1^+=-H_1^-$, the resulting operator $\Phi(x)$ will have non-zero matrix
elements only on the $2n$ particle states. Its FF are entirely expressed in
terms of the FF of the Sinh-Gordon model at $B=2/3$
\EQ
F_{2n}^{\Phi}\,=\,
\left(\mu\left(\frac{2}{3}\right)\right)^{2n} \,Q_{n}(1) \,\prod_{i<j}
\frac{F_{\rm min}^{SG}\left(\beta_{ij},\frac{2}{3}\right)}{x_i+x_j}\,\,\, ,
\EN
with ${\cal N}(B)$ and $\mu(B)$ defined in eqs.\,(\ref{integral}) and
(\ref{mu}).

Comparing the form factors of $\Box \varphi$ and the form factors of the
fundamental exponentials the quantum equation of motion can be cast in the
form
\bea
\Box \varphi(x) &\,=\,&\frac{m^2}{2\sqrt{2}}
  \sqrt{\frac{F_{\rm min}^{\rm BD}(i\pi,{\cal B})}{
\sin^2 \left(\frac{\pi}{6}({\cal B}+2)\right) \Gamma^2({\cal B})
       + 4 W({\cal B})}}
 \left(\theta(1-{\cal B})\,-\,\theta({\cal B}-1) \right)
 \left(e^{-2\lambda\varphi(x)} - e^{\lambda\varphi(x)}\right)
\nonumber \\
&\,=\,&\frac{m^2}{2\sqrt{2}}
  \sqrt{\frac{F_{\rm min}^{\rm BD}(i\pi,{\cal B})}{
\sin^2 \left(\frac{\pi}{6}({\cal B}+2)\right) \Gamma^2({\cal B})
       + 4 W({\cal B})}} \,
 \left(V_-(x,{\cal B}) - V_+(x,{\cal B}) \right)
\,\,\,.
\label{bdqem}
\eea

\subsection{Class of Stress-Energy Tensor}

The most general expression of the trace of the stress-energy tensor
compatible with the (quantum) equation of motion of the BD model can be
expressed in terms of the fundamental exponentials as
\bea
\Theta(x) &\,=\,& F_0^{\Theta}({\cal B})
\left(a\,\Phi_1(x,{\cal B}) \,+\, (1\,-\,a)\,\Phi_2(x,{\cal B})\right)
\,\,\,,
\eea
where $F_0^{\Theta}({\cal B})$ is its vacuum expectation value
\EQ
F_0^{\Theta}({\cal B}) \,=\, \frac{\pi m^2}{2 W({\cal B})} \,\,\,,
\EN
(as computed by the Thermodynamical Bethe Ansatz \cite{TBA}), whereas
$a$ is a free parameter. Varying the value of $a$, we may reach different
ultraviolet limit of the BD model. Before considering the general case, let us
analyze separately the two cases $a=1$ and $a=0$.

\subsubsection{The case $a=1$}

For this value of $a$, the trace of the stress-energy tensor is given
entirely by the operator $\Phi_1(x,{\cal B})$ and therefore we expect that
the ultraviolet behaviour will be described by a CFT with bare action given by
\EQ
{\cal S}_2\,=\,\int d^2x \left[
\frac{1}{2}(\partial_{\mu}\varphi)^2-\frac{m^2}{6\lambda^2}
\,e^{-2\lambda\varphi}\,\right]\,\,\, ,
\label{actionBD1}
\EN
and background charge \cite{HM}
\EQ
Q_2(\lambda)\,=\,-\left(\frac{\sqrt{\pi}}{\lambda}
\,+\, \frac{\lambda}{2\sqrt{\pi}}
\right) \,\,\, .
\EN
Using eq.\,(\ref{calB}), the corresponding central charge is given by
\EQ
c({\cal B})\,=\,1 + 12 Q^2_2(\lambda)
\,=\,1 + 12 \left( \frac{2-{\cal B}}{4 {\cal B}}
+ \frac{{\cal B}}{2-{\cal B}} + 1 \right)\,\,\, ,
\label{ccL-}
\EN
This is confirmed by the computation of the central charge in terms of the
$c$-theorem by using the FF of the operator $\Phi_1(x,{\cal B})$ which defines
in this case the trace of the stress-energy tensor. The result of this
computation is reported in Table 3 (see also Fig.\,2) and the sum rule
turns out to be saturated with high percentage of precision by using just the
first two FF of $\Phi_1(x,{\cal B})$.

According to the CFT (\ref{actionBD1}), the conformal dimensions of the
primary fields $e^{\alpha\varphi}$ are given by
\EQ
\Delta_2(\alpha)\,=\, - \frac{\alpha^2}{8 \pi} \,+\,
\frac{\alpha Q_2(\lambda)}{\sqrt{4 \pi}} \,\,\,.
\label{andim(-lambda)2}
\EN
and for the fundamental exponential operators of the BD model we have
\bea
\Delta_2(\Phi_2) &\,=\,& 1 \nonumber \\
\Delta_2(\Phi_1) &\,=\,& -\,\frac{1}{2}\,-\,\frac{3}
{8\pi}\,\lambda^2 \,\,\,.
\label{andimL-}
\eea
It is easy to see that these expressions are in agreement with those extracted
by looking at the short-distance behaviour of the correlators
$<T_{zz}(z,\overline z)\,\Phi_{1}(0)>$ and
$<T_{zz}(z,\overline z)\,\Phi_{2}(0)>$.
The computation are similar to that of the Sinh-Gordon model, the only
difference being the perturbative expansion
\bea
F_{\rm min}^{\rm BD}(i\pi,{\cal B}) &\,=\,&\exp\left[-8\int_0^{\infty}
\frac{dx}{x} \frac{\sinh\left(\frac{x {\cal B}}{6}\right)
\sinh\left(\frac{x}{6}(2-{\cal B})\right) \,\sinh\frac{x}{2}
\cosh\frac{x}{6}}{\sinh^2 x}
\right] \sim  \nonumber \\
&\,\sim\,&1\,-\,\left(\frac{1}{\pi}\,+\,\frac{1}{6\sqrt{3}}\right)\,
\frac{\lambda^2}{6} \,+\,o(\lambda^4)
\,\,\, ,
\eea
and therefore we will not repeat them here.

\subsubsection{The case $a=0$}

Since the trace of the stress-energy tensor is given in this case by the
operator $\Phi_2(x,{\cal B})$, the ultraviolet limit will be ruled by a CFT
with a bare action given by
\EQ
{\cal S}_1\,=\,\int d^2x \left[
\frac{1}{2}(\partial_{\mu}\varphi)^2-\frac{m^2}{3\lambda^2}
\,e^{\lambda\varphi}\,\right]\,\,\, ,
\label{actionBD2}
\EN
and a background charge given by
\EQ
Q_1(\lambda)\,=\,\left(
\frac{\sqrt{4\pi}}{\lambda} \,+\, \frac{\lambda}{2\sqrt{4 \pi}}
\right)\,\,\,.
\EN
For the corresponding value of the central charge we have
\bea
c({\cal B}) &\,=\,& 1+\,12\, Q_1^2(\lambda)
\label{ccL+}\\
& = & 1\,+\, 12 \left( \frac{2-{\cal B}}{{\cal B}}
\,+\,\frac{1}{4}
\frac{{\cal B}}{2-{\cal B}}\,+\,1 \right)\,\,\, ,\nonumber
\eea
whereas for the conformal dimension of the primary operators
$e^{\alpha\varphi}$
\EQ
\Delta_1(\alpha)\,=\, - \frac{\alpha^2}{8 \pi} \,+\,
\frac{\alpha Q_1(\lambda)}{\sqrt{4 \pi}} \,\,\,.
\label{andim(-lambda)1}.
\EN
Hence, for the fundamental exponential operators we have in this case
\bea
\Delta_1(\Phi_1) &\,=\,& 1 \nonumber \\
\Delta_1(\Phi_2) &\,=\,& -\,2\,-\,\frac{3}{4 \pi}\,
\lambda^2 \,\,\,.
\eea
These conformal data are again confirmed by the FF approach, as shown for
instance in Fig.\,2 for the central charge.

\subsubsection{The general case}

It is now easy to write down the conformal dimension $\Delta(\alpha)$ of the
exponential operator $e^{\alpha \varphi}$ and the central charge of the CFT
reached in the ultraviolet regime for generic value of the parameter $a$
appearing in the definition of the stress-energy tensor of the BD model.
The argument is similar to that already employed in the Sinh-Gordon model.
Since $\Delta(\alpha)$ is the coefficient of the most singular term obtained
in the UV-limit of the correlation function
$<T_{z z}(z,\overline z)\, e^{\alpha\varphi(0)}>$ and the FF of
$T_{zz}(z,\overline z)$ depends linearly on the parameter $a$, the conformal
dimension is given by
\begin{eqnarray}
\Delta(\alpha) & =&\, a\,\Delta_2(\alpha)\,+\,(1-a)\,\Delta_1(\alpha)\,=
\label{andimbd}\\
&=&
-\frac{\alpha ^2}{8 \pi} \,+\,
\frac{\alpha}{\sqrt{4 \pi}}\,
\left[\sqrt{\frac{2-{\cal B}}{{\cal B}}} \left(1\,-\,\frac{3}{2}\,a\right)\,+\,
\sqrt{\frac{{\cal B}}{2-{\cal B}}} \left(\frac{1}{2}\,-\,\frac{3}{2}\,a\right)
\right] \nonumber
\,\,\, ,
\eea
where $\Delta_1(\alpha)$ and $\Delta_2(\alpha)$ are given in
eq.\,(\ref{andim(-lambda)2}) and (\ref{andim(-lambda)1}) respectively.
The linear term in $\alpha$ in (\ref{andimbd}) identifies the background
charge of the corresponding Coulomb gas. Therefore the central charge of
the CFT reached in the ultraviolet limit is given by
\EQ
c\,=\,1\,+\,3\frac{({\cal B} + 6 a - 4)^2}{{\cal B}(2-{\cal B})}
\,\,\,.
\label{gencuvbd}
\EN
The check of this formula in terms of the FF approach is shown in Fig.\,3.

Observe that, with the choice
\EQ
a\,=\,\frac{4-{\cal B}}{6}
\,\,\, ,
\EN
we have identically $c=1$, a result that is confirmed by the FF approach within
the usual accuracy of few percents. The corresponding trace of the
stress-energy tensor is given by
\EQ
\Theta(x)\,=\,F_0^{\Theta}({\cal B})\,
\left(\frac{4-{\cal B}}{6}\,e^{\lambda\varphi(x)}\,+\,
\frac{2+{\cal B}}{6}\,e^{-2\lambda\varphi(x)}\right)
\,\,\, .
\EN
This operator is manifestly self-dual. In the limit $\lambda\rightarrow 0$, it
reduces to
\bea
\Theta(x) &\,=\,& \frac{2\pi m^2}{3 \lambda^2}
\left(2 e^{\lambda \varphi(x)} +  e^{- 2 \lambda \varphi(x)}\right) \,\,\, ,
\eea
which is the classical expression of $\Theta(x)$ for the Bullough-Dodd model.

\resection{Conclusions}

The ultraviolet behaviour of a two-dimensional QFT is generally characterized
by a scaling behaviour described by a CFT. The main features of a CFT are
encoded in the definition of the stress-energy tensor $T_{\mu\nu}(x)$. As
shown by the form factor approach, associated to an on-shell dynamics,
there is a one-parameter family of possible operators $T_{\mu\nu}(x)$ that
induces different scaling behaviour of the massive theory in the ultraviolet
limit. In light of their simple spectrum, we have analyzed in detail the
Sinh-Gordon and the Bullough-Dodd Models, computing the relevant CFT data
(central charge and conformal dimensions) in terms of the FF of the
fundamental exponential operators.

The ultraviolet properties of these models are strictly related to the duality
symmetry of their $S$-matrix. This symmetry has far-reaching
consequences. In fact, since the FF are computed in terms of the $S$-matrix,
this symmetry may also be extended off-shell. In particular, it gives rise
to a bidimensional representation in the space of the fundamental
exponential operators of both theories. Moreover, from the self-duality of the
two theories, we have an identification of the two operators at the self-dual
point. This remarkable property is based on our definitions (\ref{fundamental})
and (\ref{fundamentalBD}) of the fundamental exponentials of the two theories
in terms of their respective cluster operators. The validity of these formulas
has been checked by analyzing the ultraviolet behaviour of the massive
theories. An important difference between the BD and the SG models is in the
expressions of the central charge associated to their fundamental exponentials.
In fact, whereas in the SG model we have the self-dual function
(\ref{guessc1}) which is common to both the exponentials $\Phi_{\pm}(x)$, in
the BD model we have the different expressions (\ref{ccL-}) and (\ref{ccL+}),
related, however by duality
\EQ
c(a=0,{\cal B})\,=\,c(a=1,2-{\cal B}) \,\,\,.
\EN
It would be interesting to extend our results to other Affine Toda Field
Theories and to study in more detail the interplay between the massive and
conformal data of the models.

\vspace{1cm}
\noindent
{\em Acknowledgments}

\noindent
The authors wish to thank L. Bonora, S. Elitzur, E. Gava, A. Schwimmer and
Al.B. Zamolodchikov for useful discussions.

\appendix

\appsection

Solving the recursive equations (\ref{recc1}) and (\ref{recc2}), the first
FF which have the properties to be asymptotically constant and proportional
to the invariant combination $\sigma_1\sigma_{n-1}$ ($n>2$) are given by
\EQ
F_2(\be)\,=\,F_{\rm min}^{\rm BD}(\be,{\cal B})\,\left\{ H_2\,-\,
H_1\,\frac{ \sin \frac{\pi}{6}({\cal B}+2)
\,\Gamma({\cal B}) }{\sqrt{F_{\rm min}^{\rm BD}(i\pi,{\cal B})}}\,
\frac{1}{ \cosh \be \,+\frac{1}{2}} \right\} \,\,\,,
\label{ff2class3}
\EN
and
\bea
F_3(\beta_1,\beta_2,\beta_3)&=&
\left( \prod_{i<j}^3 \frac{F_{\rm min}^{\rm BD}(\be_i-\be_j,{\cal
B})}{(x_i+x_j)
(\omega \,x_i+x_j)
(\omega^{-1}\, x_i+x_j)}\right) \,
\frac{4}{F_{\rm min}^{\rm BD}(i\pi,{\cal B})}
\s_1 \s_2 \times
\nonumber \\
& & \,\,\,\,
\left\{ q_1({\cal B}) \s_3^2 \,+\,
q_2({\cal B}) \s_3 \s_2 \s_1\,-\,
\right. \label{ff3class3}\\
& & \,\,\left.
H_1\,W({\cal B})\,\left(\s_3 \s_1^3 +\s_2^3 \right)\,+\,
q_3({\cal B}) \s_2^2 \s_1^2
\right\}
\,\,\, , \nonumber
\eea
where
\bea
q_1({\cal B})&=&2 \Gamma({\cal B})\,\sin \frac{\pi}{6}(2+{\cal B})\,
\left[\cos\frac{\pi}{3}(2+{\cal B})\,-1 \right]
\left[\Gamma({\cal B}) H_1\,\sin \frac{\pi}{6}(2+{\cal B})
\,+\sqrt{F_{\rm min}^{\rm BD}(i\pi)} \frac{H_2}{2} \right]
\nonumber \\
q_2({\cal B})&=& \Gamma({\cal B}) \,\sin \frac{\pi}{6}(2+{\cal B})
\left[\Gamma({\cal B}) H_1\, \sin \frac{\pi}{6}(2+{\cal B})\,-
\sqrt{F_{\rm min}^{\rm BD}(i\pi)}H_2\,
\left(\cos\frac{\pi}{3}(2+{\cal B})\,-\frac{3}{2} \right)\,
\right]
\nonumber \\
q_3({\cal B})&=& H_1\,W({\cal B})\,-\,
\sqrt{F_{\rm min}^{\rm BD}(i\pi)} \,\frac{H_2}{2}\,
\Gamma({\cal B})\,\sin \frac{\pi}{6}(2+{\cal B})
\,\,\,.
\eea
$H_1$ and $H_2$ are two arbitrary parameters.

\newpage

\hs

\vspace{25mm}

{\bf Table Caption}

\vspace{1cm}

\begin{description}
\item [Table 1]. The first two-particle term entering the sum rule of the
$c$-theorem for the Sinh-Gordon model with the choice $a=1$ (second column)
compared with the central charge (\ref{guessc1}) of CFT.
\item [Table 2]. The first two-particle term entering the sum rule of the
$c$-theorem for the Sinh-Gordon model with the choice $a=1/2$.
It must be compared with the claimed free boson $C_{UV}=1$
UV-behaviour.
\item [Table 3]. The first two-particle term entering the sum rule of the
$c$-theorem for the Bullough-Dodd model with the choice $a=1$ (second column)
compared with the central charge (\ref{ccL-}) of CFT.
\end{description}

\newpage


\begin{table}

$$ \begin{array}{|c|c|c|}
 \hline
B
& C_{UV}^{\rm num}
& C_{UV}^{\rm Liouv}
\\
\hline
   &        &        \\
\frac{1}{10}  &
127.28994  &
127.31579 \\
\frac{1}{5}   &
67.61695 &
67.66667 \\
\frac{3}{10}   &
47.98763 &
48.05882 \\
\frac{2}{5}   &
38.40998 &
38.5  \\
\frac{1}{2}   &
32.89395 &
33.  \\
\frac{3}{5}   &
29.45222 &
29.57143 \\
\frac{7}{10}   &
27.24418 &
27.37363 \\
\frac{4}{5}   &
25.86323 &
26.  \\
\frac{9}{10}   &
25.10126 &
25.24242 \\
1   &
24.85738 &
25.  \\

& & \\ \hline
\end{array}
$$
\end{table}
\begin{center}
{\bf Table 1}
\end{center}

\newpage
\samepage
\vspace*{-1cm}

\begin{table}
$$ \begin{array}{|c|c|}
 \hline
B
& C_{UV}^{\rm num}
\\
\hline
   &        \\

\frac{1}{500} & 0.9999995 \\
\frac{1}{100} & 0.9999878 \\
\frac{1}{10} & 0.9989538  \\
\frac{3}{10} & 0.9931954  \\
\frac{2}{5} & 0.9897087  \\
\frac{1}{2} & 0.9863354  \\
\frac{2}{3} & 0.9815944  \\
\frac{7}{10} & 0.9808312  \\
\frac{4}{5} & 0.9789824  \\
1 & 0.9774634  \\
& \\ \hline
\end{array}
$$
\end{table}
\begin{center}
{\bf Table 2}
\end{center}

\newpage
\samepage
\vspace*{-1cm}

\begin{table}

$$ \begin{array}{|c|c|c|}
\hline
B
& C_{UV}^{\rm num}
& C_{UV}^{\rm Liouv}
\\
\hline
\frac{1}{10}  &
70.63001  &
70.63158  \\
\frac{1}{5}   &
41.32883  &
41.33333  \\
\frac{3}{10}   &
32.10844  &
32.11765  \\
\frac{2}{5}   &
27.98391  &
28.  \\
\frac{1}{2}   &
25.97441  &
26.  \\
\frac{3}{5}   &
25.10474  &
25.14286  \\
\frac{7}{10}   &
24.97886  &
25.03297  \\
\frac{4}{5}   &
25.42607  &
25.5  \\
\frac{9}{10}   &
26.38691  &
26.48485  \\
1   &
27.87364  &
28.  \\
\frac{11}{10}  &
29.96195  &
30.12121  \\
\frac{6}{5}  &
32.80360  &
33.  \\
\frac{13}{10}  &
36.66406  &
36.90110  \\
\frac{7}{5}  &
42.00619  &
42.28571  \\
\frac{3}{2}  &
49.67928  &
50.  \\
\frac{8}{5}  &
61.39520  &
61.75  \\
\frac{17}{10}  &
81.15833  &
81.52941  \\
\frac{9}{5}  &
120.98370  &
121.33333  \\
\frac{19}{10}  &
240.90584  &
241.15789  \\
 \hline
\end{array}
$$
\end{table}
\begin{center}
{\bf Table 3 }
\end{center}
\newpage


\newpage

\hs

\vspace{25mm}

{\bf Figure Caption}

\vspace{1cm}

\begin{description}
\item [Figure 1]. The first two-particle term entering the sum rule of the
$c$-theorem for the Sinh-Gordon model with the choice $a=1, 0$ (dots),
compared with the selfdual central charge (\ref{guessc1}) of CFT (solid line).
\item [Figure 2]. The first two-particle term entering the sum rule of the
$c$-theorem for the Bullough-Dodd model with the choice $a=1, 0$ (dots
and crosses, resp.),
compared with the non selfdual central charges (\ref{ccL-}), (\ref{ccL+})
of CFT
(solid thin line and solid thick line, resp.).
\item [Figure 3]. The first two-particle term entering the sum rule of the
$c$-theorem for the Sinh-Gordon model and the Bullough-Dodd model
with the coupling constant fixed at $B=1/2$ for different
values of $a$
(crosses and dots, resp.),
compared with the central charges (\ref{gencuv}), (\ref{gencuvbd})
of CFT
(solid thin line and solid thick line, resp.).
\end{description}

\end{document}